# A Formal Assisted Approach for Modeling and Testing Security Attacks in IoT Edge Devices

A. Bhanpurawala, K. El-Fakih, I. Zualkernan

*Abstract*— With the rapid growth in the number of IoT devices being added to the network, a major concern that arises is the security of these systems. As these devices are resource constrained, safety measures are difficult to implement on the edge. We propose a novel approach for the detection of IoT device attacks based on the use of formal modeling and mutation testing. Namely, we model the behavior of small IoT devices such as motion sensors and RFID reader as state machines with timeouts. We also model basic IoT attacks; namely, battery draining, sleep deprivation, data falsification, replay, and man in the middle attacks, as special mutants of these specifications. We also consider tests for detecting actual physical device manipulation. Mutation testing is then used to derive tests that distinguish these attacks from the original specifications. The behavior of these mutants is tested in real environment by running the tests on them. Our experiments show that derived the number of attack mutants and tests is small and thus these tests can be executed many times with limited overhead on the physical device. Consequently, our approach is not deterred by related high costs of traditional mutation testing. In addition, we also show that tests derived by our method which cover all IoT attacks do not provide good coverage of mutants derived using traditional mutation code-based operators and this indicates the need of using our method. A framework that implements our approach is presented along with some other relevant case studies.

*Index Terms*— Internet of things, edge devices, security threats, mutation testing, security faults, attacks, finite state machines with timeouts.

## I. Introduction

INTERNET of Things (IoT) systems are increasingly being deployed in smart homes [1], smart buildings [2], industrial internet of things (IIoT) [3], smart cities [4], intelligent farming and agriculture [5], smart transportation [6], supply chain management [7], and smart healthcare [8]. However, IoT systems are vulnerable to security violations [9][10]. Specifically, IoT edge devices can be hijacked (e.g., Botnet) and used to launch thousands of Distributed Denial of Service (DDoS) attacks [11]. IoT edge devices can also be involved in node capture and replay attacks [12]. It is also relatively easy to eavesdrop on IoT devices by monitoring the traffic going to and from a device [13], [14]. Furthers, sleep deprivation attacks [15] can be used to drain batteries. Default passwords while pairing [16], configuration and device authentication [17], [18], and legacy authentication mechanisms [19] can also compromise IoT devices. Even the services on a device can be exposed and lead to security violations [20]. Many IoT edge devices use mobile Apps for configuration and these Apps may suffer from overprivileged issues [21], [22] as well. Insecure hardware interfaces [23] is another key problem. Over the Air (OTA) updates for such devices can also be compromised [24], [25]. Finally, IoT edge can fall prey to side channel, spyware and backdoor pin code injection attacks [26].

A variety of techniques to enhance security of IoT systems have been proposed. These include better encryption for resource constrained IoT devices [27], software defined networks [28], and using machine learning techniques for intrusion detection [29]. One key challenge faced by the aforementioned solutions is the resource constrained nature of the simpler IoT edge devices. Solutions to resource constraint problem include off-loading the security task completely [30], using edge routers [31], or using a distributed fog approach [32]. A second challenge to proposed solutions is to ensure that the solutions work in most situations. In this context, techniques using machine learning (e.g., [33],[34],[35],[36],[37]) are inherently limited by the quality and representativeness of the data the models are trained on. Machine learning models and can over fit the data and are not necessarily generalizable [38]. Indeed, many approaches to address overfitting in machine learning models have been proposed [39], but overfitting remains an open problem. Some recent work in explainable AI (e.g., [40]) tries to address one aspect of this problem by building explainable systems to enhance trust, but do not solve the problem completely.

In this paper we propose an intrusion detection approach for the small resource-constrained IoT edge devices. The proposed approach tries to address the two problems of resource-constraints of IoT edge devices and generalization of the proposed solution. The approach is intended and limited to a wide variety of simple IoT devices including devices that collect temperature, humidity, pollution data, private entryways, infant screens, smart bulbs, proximity sensors, motion sensors, Radio-Frequency Identification (RFID) card readers, etc. The basic idea it to have an intrusion detection approach that can run on these small typically microcontroller-based edge devices efficiently with little overhead, and at the same time is systematic enough to generalize analytically. The proposed approach is novel as it is based on formal modeling of IoT devices and mutation testing. Formal modeling techniques provide mathematically rigorous approaches for the specification and validation of software and hardware systems and mutation testing deals with the generation of mutants from a given specification (or programming code) representing possible deviation from the original model (code), and then deriving tests that can distinguish each derived mutant that has a behavior different than the original specification (code). However, an impediment to mutation testing that limited its use

"This work was supported in part by AUS FRG-M-19". *(Corresponding author: K. El-Fakih).* The authors are with the College of Engineering, at the American University of Sharjah, Sharjah, Po. Box 26666, UAE (e-mails: g00081958@aus.edu, kelfakih@aus.edu, izualkernan@aus.edu).

in many applications is the high costs of deriving and detecting a large number of mutants. Accordingly, a related main line of research focused on the selection of appropriate operators that yield less mutants and tests while providing high fault coverage.

Our approach for the detection of IoT attacks focuses on modeling the behavior of small IoT edge devices as finite and extended finite state machines with timeouts. We also model common types of IoT device attacks as special mutants of these machines. Afterwards, we use mutation testing for deriving tests that detect or kill such mutants; i.e., distinguish the attacks from the normal behavior of the specification machine. In addition, we present a framework for implementing our method and provide relevant case studies. More precisely, the contributions of this work can be summarized as follows:

- Model the behavior of small IoT edge devices such as motion sensor, ultrasonic motion sensor, and the RFID card reader as state machines with timeouts.
- Consider common types of IoT device attacks; namely, the basic battery draining, sleep deprivation, data falsification, replay, and man in the middle attacks, and show how each of these attacks can be modeled as a collection of special mutants of the corresponding timed state machine specifications. Then, we use the mutants and their specifications to drive tests that cover these attacks. We also consider tests for detecting actual physical device manipulation.
- Build an environment that incorporates the proposed work and assess the work using this environment. The environment includes related hardware architecture, an IoT framework for running a program on microcontroller, data collection for observing the behavior of the program and a testing framework for the derivation and injection of mutants/faults into the program and for detection of these faults in practice.
- We assess the feasibility of applying the work in practice; in addition, we assess the impact of threats on battery drainage, determine the costs of running the tests on battery drainage, and determine the coverage of security tests against traditional code-based faults showing the need for considering such tests.

The rest of the paper is organized into the following sections. An introduction on timed state machines, IoT device architecture, mutation testing. We then provide state machine models for the IoT devices (Motion sensor, Ultrasonic motion sensor, and RFID). This is followed by an introduction to common security threats in IoT devices. Then, we show how each of these threats can be modeled a special mutant of the state machine models. Then we present the overall architecture of how (mutation) testing is implemented for the entire system. We then provide four case studies with some results to support our proposal.

## II. BACKGROUND

### A. Timed state machines

A Finite State Machine (FSM) model of a system is specified as finite sets of states, inputs, outputs, and transitions between states each labeled by an input/output interaction. The Extended FSM (EFSM) model extends the FSM model with parameterized inputs/outputs, variables, predicates, and update statements. Thus, a transition can be labeled by a pair of possibly parameterized input/output interactions and the transition can be guarded by a predicate (guard) that must hold true for the transition to be executed. Upon its execution, new values can be assigned to variables based on the transition update statements. Both these models can be extended with a clock to represent the inclusion of time aspects. The clock specifies the timeout at the state and it can be used for limiting the stay at the state. This means that if no input is applied at the current state before the timeout expires then the timeout is executed and the machine resets the timer while moving to the target state. State machines are as the underlying models of industrial strength tools such as SDL and UML state charts, etc. In the following we define these models as we use them to precisely describe the behavior IoT devices and their attacks. For more information about state machine models and mutation testing approached and applications the reader may refer to Mathur [41] for foundation of testing, Merayo et. al [42] for extend timed FSM models, Bresolin et al. [43] for dealing with deterministic timed FSM models, and Bresolin et al. [45] for equivalence checking of timed FSMs.

An FSM is a 5-tuple $(S, I, O, h_S, s_1)$ where $I$ and $O$ are input and output alphabets, $S$ is a finite non-empty set of states with the designated initial state $s_0$, and $h_S \subseteq (S \times I \times O \times S)$ is the transition relation. A transition in $h_S$, given in the form $(s, i, o, s')$ indicates that if the machine is at current state $s$, upon receiving the input $i$, it produces the output $o$ while moving to state $s'$. An *FSM with timeouts*, a timed FSM (TFSM) for short, is an FSM annotated with a *clock* that is reset to 0 at the execution of any transition. Such a TFSM can have input timeout transitions. When an input timeout expires at a state, the TFSM can spontaneously move to the destination state of the timeout transition while resetting the time to 0. Thus, a TFSM is a 6-tuple $Q = (I, S, O, h_S, \Delta_S, s_1)$ where $\Delta_S: S \rightarrow S \times (N \cup \{\infty\})$ is the timeout function, where $N$ is the set of positive integers: for each state, this function specifies the maximum time for waiting for an input. Given state $s$ of $Q$ such that $\Delta(s) = (s', t_{out})$, if no input is applied before $t_{out}$ expires, then at $t_{out}$, $Q$ moves to state $s'$ and the clock is set to 0. If $s = s'$ then the clock is set to zero when the timeout is expired. A transition $(s, i, o, s') \in h_S$ means that $Q$ being at state $s$ accepts an input $i$ applied at time $t < t_{out}$, measured from the moment when the clock was reset at state $s$, then $Q$ produces $o$ and the clock then set to 0.

Given a TFSM $Q$, a *timed input* is a pair $(i, t)$ where $i \in I$ and $t$ is a real; a timed input $(i, t)$ means that input $i$ is applied to $Q$ at time instance $t$ where $t$ is a local time (at the state). A timed input of a timeout transition is written as $(\varepsilon, t_{out})$ and its output is $\varepsilon$, where $\varepsilon$ is the empty string. A sequence of timed inputs $\alpha = (i_1, t_1)(\varepsilon, t_{out}) \ldots (i_n, t_n)$ is a *timed input sequence*. A sequence α/γ = ($i_1$, $t_1$)/$o_1$ (ε, $t_{out}$)/ε … ($i_n$, $t_n$)/$o_n$ of consecutive pairs of timed inputs and outputs starting at the state $s$ is a (*timed*) *trace* of $Q$ at state $s$. A trace of $Q$ starts from the initial state $s_0$. A *test case* of TFSM $M$ is a finite length trace of $M$. A *test suite* is finite set of test cases. Note for simplicity of presentations, in the figures, we use $s$ - $t_{out}$ → $s'$ to denote the timeout transition from $s$ which leads at $t_{out}$ to state $s'$.

As a simple example, consider the TFSM of the motion sensor in Fig.3. Initially, at the initial state $S_1$, the clock starts from 0, then at time $t_{out}$ = 2 it moves to state $S_2$ while resetting the clock to 0. Then at $S_2$ if $i_1$ is not provide at time $t < t_{out}$ = 2, then at $t = 2$, $t_{out}$ is executed and the machine moves back to $S_1$ while resetting the clock to 0. However, if at $S_2$, $i_1$ is provided at time $t < 2$, then the machine executed transition $t_4$ where it produces $o_1$ and moves to state $s_3$ while resetting the clock to 0. The traces that correspond to these two scenarios are (ε, $t_{out}$ = 2) (ε, $t_{out}$ = 2), and (ε, $t_{out}$ = 2) ( ($i1$, $t < 2$ ), $o_1$), respectively.

In this paper, we also use the extended (timed) FSM model for modeling IoT devices with variables and updates. The extended FSM (EFSM) model extends the FSM model with variables $V$, update statements, guards, and parameterized inputs and outputs. Thus, an EFSM $M$ has also finite sets of variables $V$ and transitions $T$. Let us assume that there are $n$ variables. $V$ is a set of $n$ variables and $v^0$ is the vector of initial values of these variables. We write v = ($v_1$, $v_2$, … $v_n$) for an assignment of values to the $n$ variables of $M$. The initial value assignments of the variables is called the *initial valuation* $v^0$. A pair ($s$, v) is called a *configuration* and ($s$, $v^0$) is the initial configuration of the machine representing the initial values of variables.

A transition $T_j \in T$ has the form $T_j$ = ($s$, $i(p)$, [$G$], *up*, *o*, $s'$ ),

where $s$ and $s'$ are the *current and ending* states of $T_j$, $i \in I$ is an input interaction with the parameter $p$, [$G$] is the *enabling guard* (or *predicate*) of $T_j$ which depends on the values of the state variables and the value ₽ of parameter $p$, and *up* is a concurrent update statement which defines new values for certain variables in $V$ as functions of the current values of all variables of $M$ and ₽. An update statement of variable $v$ in $V$ is written in the form { $v$ := *expr* ; }, where *expr* is an expression over the variables in $V$ and the parameter $p$. The EFSMs presented in this paper use interactions with single parameters. Note an input $i$ may not parameterized; thus, written as $i$.

The meaning of $T_j$ is the following: If $M$ is in state $s$, then $M$ may make a transition to state $s'$ by receiving the input $i$ with parameter value ₽ if [$G$] is True for the current values v of the variables and ₽. If the transition is executed, the values of v will be changed to v' according to *up*, and afterward the output *o* will be produced. An EFSM *with timeouts*, a timed EFSM for short, is an EFSM with a single clock that is reset at the execution of each transition. In addition, at each state the machine has a timeout function $\Delta_S$ similar to TFSMs.

Given a timed EFSM $N$, a *timed input* is a pair ( $i($ ₽ $)$, $t$) where $i \in I$, ₽ is a value of parameter $p$, and $t$ is a real; a timed input ( $i($ ₽ $)$, $t$) means that input $i$ carrying the parameter value ₽ is applied to the machine at time instance $t$ where $t$ is a local time. A timed input of a timeout transition is a defined above for TFSMs. A sequence of timed inputs α = ( $i_1$(₽$_1$), $t_1$) (ε, $t_{out}$)… ($i_n$, $t_n$) is a *timed input sequence*. A sequence α/γ = ( $i_1($ ₽ $)$, $t_1$)/$o_1$ (ε, $t_{out}$)/ε … ($i_n$, $t_n$)/$o_n$ of consecutive pairs of timed inputs and outputs starting at the initial configuration is a *timed trace* of timed EFSM at state $s$. A *test case* of timed EFSM $M$ is a finite length trace of $M$. A *test suite* is finite set of test cases. For detailed formal definitions of timed FSM and EFSM models, the reader may refer to [42][44].

As an example, consider the timed EFSM of the RFID reader in Fig. 5 with variables *counter* (integer) *rchar* (char with possible values [*a*---*z*, 0---9]). At $S_{11}$, assume *counter* = 0 and the string *code* is empty (ε), if the input *data_byte*('B'); i.e., *rchar* = 'B', is received at time $t_1$, as the guard [*counter*] < 10 of $t_{23}$ holds, then $t_{23}$ = ($S_{11}$, *byte_code*(*rchar*), [*counter* < 10 ], { *counter* := *counter* +1 ; *code* := *code* + *rchar* }, $o_{13}$, $S_{11}$) can execute, where counter is incremented by one (*counter* := *counter* +1) and the received character 'B' is appended to code, *code* := *code* + *rchar*, then the output $o_{13}$ is produced, and the machine moves to state $S_{11}$ again while setting clock to 0. Thus, in fact the machine moves from configuration ($S_{11}$, 0, ε) to ($S_{11}$, 1, "B"). Then, if *data_byte*('A') is received at time $t_2$ then $t_{23}$ is executed again and the machine moves to ($S_{11}$, 2, "BA"). This is realized by the timed trace (*data_byte*('B'), $t_1$))/$o_{13}$ (*data_byte*('A'), $t_2$))/$o_{13}$.

### B. Model and Code Based Mutants
#### 1) Model Based Mutants

Given a state machine (timed FSM or timed EFSM) $N$ representing the behavior of an IoT device, we model the behavior of typical device attack as a mutant of $N$. Then, we use the derived mutant and $N$ to derive test that *detects/kills/distinguishes* such attacks from $N$. A test case *distinguishes* $N$ from a mutant if the output sequences of $N$ and the mutant, with respect to the input sequence of the test case, are different. Approaches for deriving such tests are elaborated in many books and papers including [44][46][47][48]. In particular, the following types of state machine mutants [46][47] are considered in this paper.

A mutant of $N$ has a *single output fault* if it has a transition with an output different than that specified at the corresponding transition of $N$. That is, for certain transition of $N$ with output $o$, the corresponding transition in the mutant has an output $o' \neq o$, $o \in O$.

In addition, here we define the following types of mutants: A mutant has a *timeout* fault if it has a state with a different timeout than that specified at the corresponding state in machine $N$; i.e. for certain $s$ of $Q$ with timeout $t_{out}$, state $s$ in the mutant has timeout $t'_{out}$ such that $t'_{out} \neq t_{out}$. A mutant has an *added transition fault* if the mutant is obtained by adding a new transition to $N$. A mutant has a *new state fault* if it is obtained from $N$ by adding a new state with related outgoing transitions and modifying the ending states of some other transition(s) such the added state becomes reachable through this (these) transitions.

#### 2) Code-based mutants and mutation score

We present C++ implementations of the considered IoT devices. In fact, we consider testing the attacks in a real system.

To this end, for each considered attack, we inject the attack into the code implementation creating a code mutant of that implementation. Then, we run the corresponding test case on the mutant to check if such an attack is detected in practice. Furthermore, we create mutants from the implementation considering traditional C based mutation operators illustrated in [41] to assess the effectiveness of derived (attack) test suites in detecting traditional (none attack based) code faults. The mutation score MS of a certain test suite [41], with respect to a given collection of mutants, is computed based on the percentage of the number of mutants killed by the suite divided by the number of all derived mutants minus the number of alive mutants that cannot be killed by any test suite. Alive mutants are determined by running a collection of many test suites against the mutants and those which not killed by any suite are considered alive. Thus a test suite that kills all non-alive mutants has a 100% mutation score.

### III. MODELING IoT DEVICES AS TIMED FINITE STATE MACHINES

In this section we show how simple IoT devices, such as the motion sensor, ultrasonic motion sensor, and the RFID card reader, can be modeled using timed FSMs/EFSMs.

### A. Motion sensor

#### 1) Description

For a Passive Infrared motion sensor (PIR) [49] is a classic example of a sensor often used in IoT scenarios especially in the context of smart homes. A PIR sensor measures the infrared light radiating from objects or human bodies nearby to detect whether the user is approaching or not [50]. Fig. 1 shows a typical IoT architecture for a motion sensor. The motion sensor is connected to a microcontroller that in turn is using Wi-Fi to connect to a home router which is connected to the Internet. Up detecting motion, the microcontroller uses the MQTT protocol to send a message to the Application server and potentially to Apps running on a consumer's laptop or a mobile phone. From a communication perspective, the microcontroller first must connect to the WiFi network and then connect to a MQTT Broker using TCP. A microcontroller used in such a scenario will typically not be multi-threaded and run in a single loop. Once connected to WiFi and the MQTT broker, the microcontroller continuously checks for a movement signal from the PIR motion detector. Once a signal is detected debouncing is done by sleeping for a few milliseconds. This ensures that multiple messages are not generated for the same movement event. For each movement event, an MQTT message is then published.

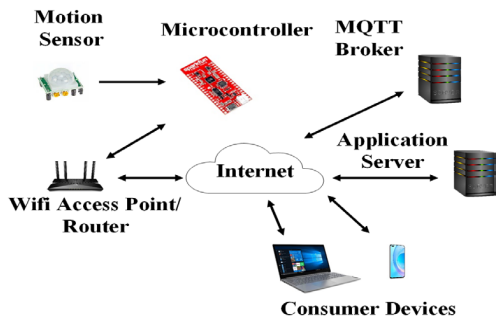

**Fig. 1.** An IoT Architecture for the PIR motion detector.

#### 2) Motion Sensor Timed FSM Model

The working of the motion sensor has been modelled as timed FSM in Fig. 2. Fig. 2 is essentially a representation of the C++ program running on the microcontroller. Table I shows the various inputs and outputs of the state machine. As Table I shows, inputs are events like health check ($i_5$) from the application server or physical events like motion status ($i_{11}$). Outputs constitute a console output or an MQTT messages that is sent to the MQTT broker. For example, $o_5$ represents an output or an MQTT message indicating that the sensor is alive. The program is single threaded and runs in a continuous loop. As Fig. 3 shows, upon powering up, the microcontroller first tries to connect to the WiFi ($S_2$) and upon connecting issues the "Connected to wifi" is printed to console and goes into the connected state ($S_3$). There are timeouts for each state to ensure that the microcontroller does not get hung up in any state. For example, failing to connect to WiFi, the microcontroller goes back to the dwell state ($S_1$). Once the WiFi connection is established, the microcontroller moves to the connection to MQTT state ($S_4$) to try to connect to the MQTT broker. Once connected, the microcontroller moves between states $S_5$ and $S_{10}$. This is because it is single threaded, the microcontroller needs to check for any incoming messages from the MQTT broker and also to periodically check for motion. While in $S_5$, many input events like $i_5$ = "Health Check" or $i_6$ = "Unexpected Request" are received and responded to with appropriate messages (e.g., $o_5$ = "Sensor is Alive").

Once the motion is detected in $S_{10}$ (i.e., $i_{11}$ = Motionstatus(1)), the microcontroller moves into the dwell state ($S_{11}$) for debouncing the signal after issuing $o_{10}$ = "Motion has been detected" message.

### B. Ultrasonic motion sensor

#### 1) Description

The ultrasonic motion sensor is an electronic device that makes use of the SONAR technique to detect the distance of any object around it. It comprises of 2 ultrasonic transmitters which are a receiver and a control circuit. The transmitter emits a sound with high ultrasonic frequency. The echo produced from this sound is recorded by the receiver to calculate the time difference between the sound emitted and the echo received. HC-SR04 ultrasonic motion sensor was used for this for research. The architecture for a ultrasonic motion sensor is similar to that of a PIR sensor except that instead of a signal for movement a number representing the estimated distance is received by the microcontroller.

#### 2) Ultrasonic motion sensor Timed Extended FSM Model

Fig. 4 shows the TEFSM for the ultrasonic motion sensor. States $S_1$ to $S_9$ of the ultrasonic motion sensor are similar to those of the motion sensor given in Fig. 3 with the same inputs $i_1$ to $i_9$ and $i_{11}$ and outputs $o_1$ to $o_9$. However, the ultrasonic has different $S_{10}$, $S_{11}$ and $S_{12}$ that are shown in Fig. 4. When the sensor goes into the active read state ($S_{10}$), it reads in transition $t_{21}$ the value from the echo pin through the parameterized input $i_{11}$ = echoPin(duration) which receives the (integer) parameter duration; then based on that it calculates (integer variable) distance using the update *distance* := (duration * 0.034)/2; and the outputs $o_{10}$ = "duration" and moves to state $S_{11}$. Afterwards, using input and $i_{12}$ = DistanceStatus(1), the sensor outputs $o_{12}$ = "distance" and moves to the dwell state $S_{12}$.

TABLE I
DESCRIPTION OF THE INPUTS AND OUTPUTS OF THE MOTION SENSOR

| Input | | Value | Description |
|---|---|---|---|
| $i_1$ | = | ws(0) | indicates that Wifi is not connected and trying to connect |
| $i_2$ | = | ws(1) | indicates that WiFi is connected |
| $i_3$ | = | cc(0) | indicates that MQTT queue (client) not connected and trying to connect |
| $i_4$ | = | cc(1) | indicates that MQTT queue (client) is connected |
| $i_5$ | = | "Health Check" | |
| $i_6$ | = | "Unexpected Request" | |
| $i_7$ | = | "Stop Publishing" | |
| $i_8$ | = | "Give me wifi Name" | |
| $i_9$ | = | buttonPress(low) | |
| $i_{10}$ | = | "Start Publishing" | |
| $i_{11}$ | = | Motionstatus(1) | |

| Output name | | Description |
|---|---|---|
| $o_1$ | = | "Connecting to wifi" |
| $o_2$ | = | "Connected to wifi" |
| $o_3$ | = | "Connecting to MQTT" |
| $o_4$ | = | "Connected to MQTT" |
| $o_5$ | = | "Sensor is Alive" |
| $o_6$ | = | "Error. Invalid message" |
| $o_7$ | = | "Publishing has stopped" |
| $o_8$ | = | "Press any button" |
| $o_9$ | = | "Publishing has started again" |

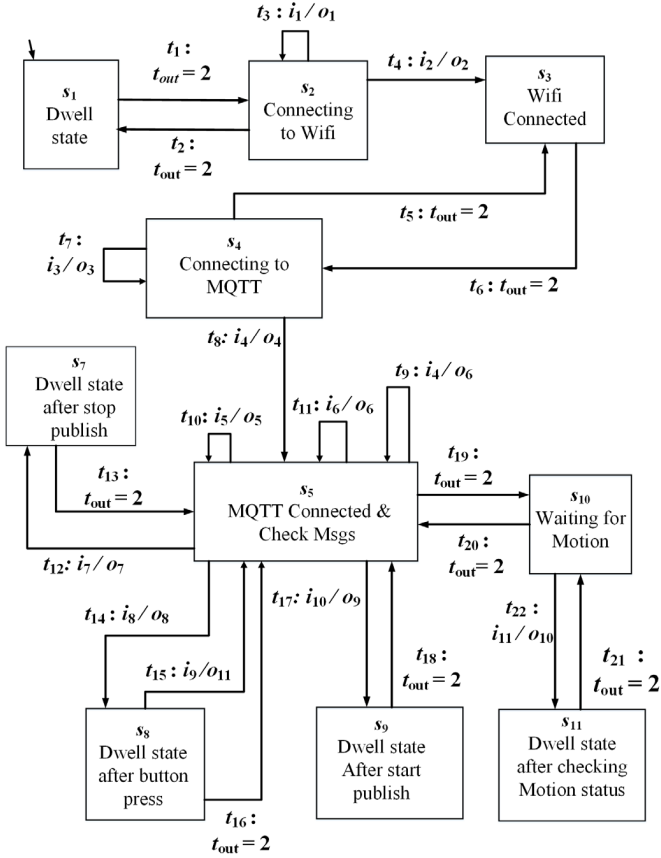

**Fig. 2.** Timed FSM of the motion sensor.

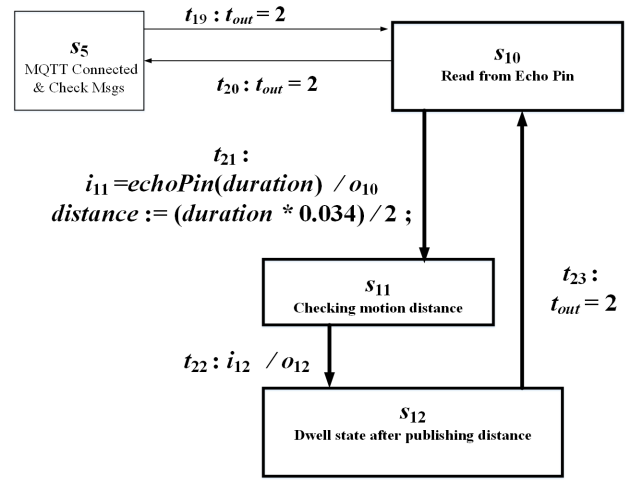

**Fig. 3**. Part of the timed extended FSM of the Ultrasonic Motion Sensor. The other part is similar to the motion sensor in Fig. 2.

C. RFID Card Reader

1) Description

Radio-Frequency identification (RFID) is often used in IoT applications for identification [51]. In this interaction when an RFID tag comes near an RFID card reader magnetic field is used for communications and the tag's unique ID (typically 10 digits) is sent to the microcontroller. The architecture of an RFID card reader scenario is very similar to the one shown in Fig. 2. The only difference is that the PIR sensor is replaced by an RFID card reader and the microcontroller receives 10 digits instead of signal for motion

2) *RFID Card Reader timed extended FSM model*

The working of the RFID card reader has been modelled as a timed extended Finite state machine as shown in Fig. 5. Table II shows all the inputs and outputs in the state diagram. The state diagram assumes that the 'valid' number for the card is known in advance and hence represents a constant for the state diagram. This constant is used to determine if the 10-digit number received is a valid card or not. Most of the state diagram is similar to that for PIR shown in Fig. 2.

**Fig. 4.** Timed EFSM of the RFID.

Table II shows the inputs and outputs of the RFID reader. The primary difference is reflected in states $S_{10}$, $S_{11}$, and $S_{12}$. The input $i_{13}$ = "$0A" triggers the start of reading the 10 digits from an RFID card. Subsequent characters are read in state $S_{11}$ and when the counter reaches 10 [$counter = 10$], the microcontroller goes into state $S_{12}$ to wait for the stop byte. Depending on if the card is acceptable [$code = card$], the microcontroller moves back to state $S_{10}$ with different messages after resetting *counter* to 0.

TABLE II
DESCRIPTION OF THE INPUTS AND OUTPUTS OF THE RFID CARD READER

*Variables:*
  code(string),
  card = "0003490808" (a constants of type string representing the card number) – In reality it will be variable, but for this purpose we are keeping it constant (preprogrammed to only accept one card)
  counter (integer),
*Parameterized input:*
$i_{11}$ = *data_byte*(*rchar*) with parameter *rchar* of type String

| Input | | Value | Description |
|---|---|---|---|
| $i_1$ | = | ws(0) | indicates that Wifi is not connected and trying to connect |
| $i_2$ | = | ws(1) | indicates that WiFi is connected |
| $i_3$ | = | cc(0) | indicates that MQTT queue (client) not connected and trying to connect |
| $i_4$ | = | cc(1) | indicates that MQTT queue (client) is connected |
| $i_5$ | = | "Health Check" | |
| $i_6$ | = | "Unexpected Request" | |
| $i_7$ | = | "Stop Publishing" | |
| $i_8$ | = | "Give me wifi Name" | |
| $i_9$ | = | buttonPress(low) | |
| $i_{10}$ | = | "Start Publishing" | |
| $i_{12}$ | = | "$0D" | |

| Output name | | Value |
|---|---|---|
| $o_1$ | = | "Connecting to wifi" |
| $o_2$ | = | "Connected to wifi" |
| $o_3$ | = | "Connecting to MQTT" |
| $o_4$ | = | "Connected to MQTT" |
| $o_5$ | = | "Sensor is Alive" |
| $o_6$ | = | "Error. Invalid message" |
| $o_7$ | = | "Publishing has stopped" |
| $o_8$ | = | "Press any button" |
| $o_9$ | = | "Publishing has started again" |
| $o_{10}$ | = | "Start byte received" |
| $o_{11}$ | = | device_ssid |
| $o_{12}$ | = | "Request coming from invalid IP" |
| $o_{13}$ | = | "Card is reading data bytes" |
| $o_{14}$ | = | "Card rejected" |
| $o_{15}$ | = | "Card accepted" |
| $o_{15}$ | = | "Null" |

## IV. IOT ATTACKS

### A. Security Issues in IoT devices

Table III summarizes some common security issues in IoT devices addressed in this paper [15] [52]. The modelling of each attack is described below.

 

TABLE III
TYPES OF ATTACKS IN IOT DEVICES

| Input | Type of Attack | Description |
|---|---|---|
| $A_1$ | Battery Draining | Forcing the edge device to execute power-consuming subroutines. |
| $A_2$ | Sleep Deprivation | Preventing the node from going to sleep. |
| $A_3$ | Data falsification | In this type of attack the information known about a systems operation allows the attacker to falsify data by planting malicious nodes or turning a normal node into a malicious node |
| $A_4$ | Replay attack | A Replay Attack is made by spoofing, altering, or replaying the identity information of smart devices in the IoT network |
| $A_5$ | Man in the middle | In this attack, an attacker or hacker intercepts a communication between two systems. |

### B. Attacks modeled as state machine mutants

In previous sections, we modeled the behavior of the motion sensor, ultrasonic motion sensor, and RFID card reader as timed state machines and introduced common types of IoT device attacks. In this section, we show how these attacks can be represented/modeled as mutants of the considered state machines. This allows the derivation of tests that detect such attacks.

#### 1) Battery Draining attack

At some states, decreasing the value of $t_{out}$ forces the machine to leave the state early decreasing the possibility of providing the user with a desired output; i.e., forcing the user to resend the intended request again and again. This causes battery drain. In our case, mutants with such anomaly can be derived by decreasing at appropriate states the $t_{out}$ period. In particular, for the motion sensor, we consider states $S_5$, $S_7$, $S_8$, $S_9$, and $S_{10}$, then for each state, the timeout period of the state is decreased from 2 to 1 second, creating a corresponding mutant with timeout fault. Similarly, for the ultrasonic, the timeout period at states $S_5$, $S_7$, $S_8$, $S_9$, $S_{10}$, $S_{11}$, and $S_{12}$ are changed in creating the collection of related mutants, and for RFID timeouts at states $S_5$, $S_7$, $S_8$, $S_9$, and $S_{10}$, are changed creating the related collection. In total, 5, 7, and 8 mutants are created for the motion sensor, ultrasonic sensor, and the RFID, respectively.

#### 2) Sleep Deprivation attack

This type of attack can be simulated by a mutant that prevents the device from sleeping at certain states. For instance, in the motion sensor (Fig. 4), at state $S_{10}$ (Waiting for motion ) the device wakes up after every 2 seconds to check if there is motion and then goes back to sleep. For ultrasonic motion sensor it happens at the state $S_{10}$ and for the RFID reader this can happen at state $S_{10}$ (Ready to Read). In each case, we derive the corresponding mutant by reducing the timeout at these states from 2 to 0.001 seconds. Though the reduction of the tout time can be done at every state; yet, the above selected states are the most relevant to sleep deprivation attacks.

#### 3) Data falsification attack

This attack can be simulated by changing the output of certain transition at certain state (i.e. by creating a mutant with an output fault). For example, if the sensor is supposed to reply (output) the health status, it would reply with an error message instead. For the motion sensor, this can happen by replacing the output to $o_6$ by another output at transitions $t_7$, $t_8$, $t_{12}$, $t_{14}$, $t_{15}$, $t_{17}$, and $t_{22}$. For each replaced output, the corresponding mutant is created. Similarly, for the ultrasonic, this can happen by changing the output to $o_6$ at transitions $t_7$, $t_8$, $t_{12}$, $t_{14}$, $t_{15}$, $t_{17}$, $t_{21}$, and $t_{22}$. For the RFID, output $o_6$ is changed at transitions $t_7$, $t_8$, $t_{12}$, $t_{14}$, $t_{15}$, $t_{17}$, $t_{22}$, $t_{23}$, $t_{24}$, $t_{25}$, $t_{26}$, and the corresponding mutants are included in a related. In total, 7, 8, and 11 mutants are created considering for the motion sensor, ultrasonic sensor, and the RFID, respectively.

#### 4) Replay attack

In the motion sensor (Fig. 3), input $i_4$ at transition $t_8$ takes the machine from state $S_4$ (Connecting to MQTT) to $S_5$ (Connected to MQTT) while providing the output $o_4$ (Connected to MQTT). This indicates that the connection has been established; and then afterwards, if the input $i_4$ is provided at state $S_5$, the machine responds with the output $o_6$. However, once the connection is established, an attacker can simulate again at state $S_5$ the behavior of $t_8$; this can be done by creating a mutant by adding as a self-loop transition at $S_5$ with the same input and output as that of $t_8$. The same can be replicated for the ultrasonic motion sensor and RFID card reader at the same state $S_5$.

#### 5) Man in the middle attack

For the motion sensor, as shown in Fig. 6, at state $S_{10}$ (Waiting for motion), if the input $i_{11}$ (Motionstatus(1)) is applied the machine outputs $o_{10}$ (Motion has been detected), and moves to a Dwell state ($s_{11}$). To model man in the middle, a new state $S_{new}$ is added such that $i_{11}$ takes the machine from $S_{10}$ to $S_{new}$ instead of the Dwell state $S_{11}$. A timeout transition $t_{out} = 2$ is also added connecting $S_{new}$ back to $S_{10}$ again; thus, preventing the user to reach $S_{11}$ if the user does not provide within 2 seconds the input $i_{11}$ again. $S_{new}$ takes this information and passed it on to $S_{11}$. This type of attack can happen at many states; however, we have kept the most relevant states for this type of attack as knowing the status of the motion in a sensor would be the most valuable information that an attacker would have. However, this $S_{new}$ state can be easily replicated between states $S_5$ and $S_7$, $S_5$ and $S_8$, $S_5$ and $S_9$ for all 3 devices: the motion sensor, the RFID card reader and the ultrasonic motion sensor as well.

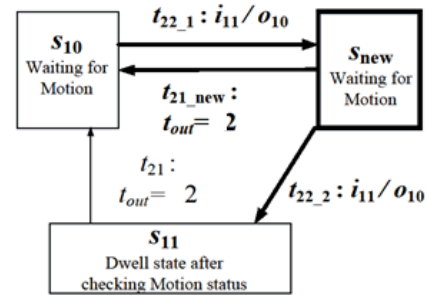

Fig. 5. Man in the middle attack for a motion sensor.

### 6) Mutants with increased timeout for both battery draining and sleep deprivation

At some states, we increased the value of $t_{out}$ such that the device spends more time in performing the operations and also would let to spend more time in the dwell state. This causes the device to spend more time in each of the states. Mutants with such anomaly can be derived by increasing the $t_{out}$ period at appropriate states. In particular, for the motion sensor, we consider states $S_5$, $S_7$, $S_8$, $S_9$, and $S_{10}$, then for each state, the timeout period of the state is increased from 2 seconds to 5 seconds, creating a corresponding mutant with timeout fault. Similarly, for the ultrasonic, the timeout period at states $S_5$, $S_7$, $S_8$, $S_9$, $S_{10}$, $S_{11}$, and $S_{12}$ are changed in creating related collection of mutants, and for RFID timeouts at states $S_5$, $S_7$, $S_8$, $S_9$, and $S_{10}$ are changed for creating related mutants. In total, 5, 7, and 8 mutants are created for the motion sensor, ultrasonic sensor, and the RFID, respectively.

### C. Physical Device Attacks

One of the critical issues about edge devices is that they can be physically tampered. There is a need detect if the device has been compromised physically. For example, the attacker can just remove the edge device and replace it with its own or the attacker can just disconnect the device from the system so that it stops sending signal. First, the code is tested against the mutants behavior and checked that it works well and thus can detect all the mutants. Then, we replicated the working of a motion sensor on a second ESP which is connected to the first ESP. Whenever want to send an input like one from the motion sensor, we simulate it from the second ESP instead. Once it is in complete working state, we inject (the considered attack) faults into this replica and analyze this behavior. Since we have complete control on the simulation, we can look at a step by step analysis of where exactly the system fails if a fault is detected. It is not just the regular behavior of the replica that can help us in analysis, we also analyze the behavior when there are faults introduced. Thus, if the replica does not respond similar to the motion sensor to the test cases; we conclude that this replica is not the original sensor and thus it might have been physically tampered and replaced**.** Also, another way to check for a physical device compromise is by injecting faults on a compromised device and compare if it the abnormal behavior of the compromised device is the same as the abnormal behavior of a replica. The comparison of these two can give us more insights and help us conclude if the device has been physically changed. At some states, we increased the value of $t_{out}$ such that the device spends more time in performing the operations and also would let to spend more time in the dwell state. This causes the device to spend more time in each of the states. Mutants with such anomaly can be derived by increasing the $t_{out}$ period at appropriate states. In particular, for the motion sensor, we consider states $S_5$, $S_7$, $S_8$, $S_9$, and $S_{10}$, then for each state, the timeout period of the state is increased from 2 seconds to 5 seconds, creating a corresponding mutant with timeout fault. Similarly, for the ultrasonic, the timeout period at states $S_5$, $S_7$, $S_8$, $S_9$, $S_{10}$, $S_{11}$, and $S_{12}$ are changed in creating related collection of mutants, and for RFID timeouts at states $S_5$, $S_7$, $S_8$,

$S_9$, and $S_{10}$ are changed for creating related mutants. In total, 5, 7, and 8 mutants are created for the motion sensor, ultrasonic sensor, and the RFID, respectively

## V. IoT Testing framework

### A. Experimental Setup

Fig. 6 shows the experimental setup including various components and how they communicated with each other. As show in Fig. 6, the sensors/actuators (motion sensor or an RFID card reader) are connected to the edge device (ESP32) which in turn is connected to the MQTT broker using WiFi. The messages sent and received from the broker can be viewed by the Middleware which was implemented using Node-Red. CloudMQTT was used for creating the MQTT queue the messages are monitored using Node-Red with MQTT QoS level 2. The edge devices have been programmed in a way that to explicitly model the FSMs that are shown in the above sections. When the device is running, an input is given via the Node-red interface, which is same as an input given to a state of the FSM.

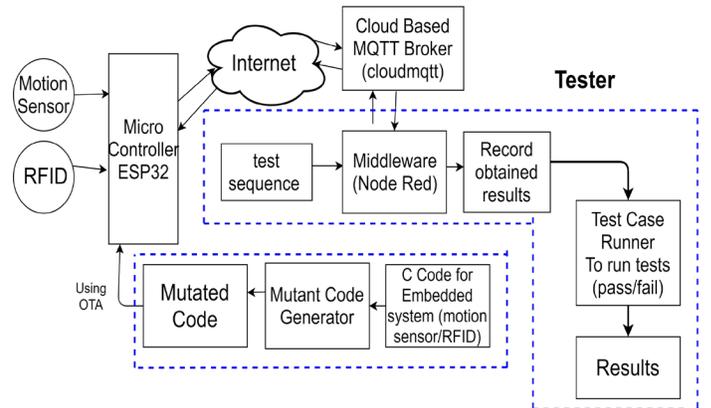

**Fig. 6.** Experimental Setup.

### 1) The Edge Device

The sensors/actuators used for this experiment were PIR motion sensor and Parallax RFID Card Reader. We used the *The SparkFun ESP32 Thing* Microcontroller as the edge device because of its in-built WiFi and OTA capability. The hardware specification for ESP32 is given in Table IV.

TABLE IV
HARDWARE SPECIFICATION FOR THE EDGE DEVICE

| Specification | SparkFun ESP32 Thing |
|---|---|
| Processor | Dual-core Tensilica LX6 microprocessor |
| SRAM | 520kB |
| Networking | 802.11 b/g/n/e/i |
| Storage | 4MB Flash memory |
| I/O | 28 GPIO |
| On-board Peripherals | LED PWM |

### 2) Data collection of the device behavior

After the input is sent to the edge device using an MQTT message, the edge device responds with an output using MQTT, which is eventually recorded using the Node-red Middleware. The outputs have also been programmed in a way that corresponds to the states of the FSM. For example, if the input given to the motion sensor is asking for health check, the message is sent via MQTT on the Node-red interface to the sensor, the sensor's response that it is alive is also recorded using MQTT and the result is displayed on Node-red. The behavior and the input/output sequences are then recorded to and stored into a database for further analysis

*3) Injecting the mutants into the microcontroller*

### a) Mutants Generator

We generate the mutants that must be injected into the microcontroller in two different ways. The first type of mutants are the code mutants that are generated by creating variations of the C++ code written for the working of the edge devices. The mutant code generator takes the C++ code files and creates mutated code files which contains one mutant per file. This was implemented using Python and C++ that runs a program to scan the C++ code and then programmatically tweak/change the input file (a variable change, alter an arithmetic operator, comment a line, etc.). The mutated file is regenerated and saved in a folder. This program runs several times based on the possibilities found for code alteration and creates a separate mutated file with one mutant in each. The security mutants were created manually, by changing the code at some specific places that correspond to the security attacks mentioned in the above sections.

### b) OTA Module

While the system is running under the normal circumstances, the attacks are done in real time. Hence to keep our testing scenarios as real as possible, we wanted to inject the mutants while the system is up and running. This was done using the Over the Air (OTA) Transfer functionality provided by Arduino. While the system is running using the original code, we take the mutated C file and inject it into the hardware using the OTA functionality. Using this method, the mutated program file is uploaded to the edge device using a WiFi connection instead of serial port. So, while the system is still running, the mutated file is uploaded and accepted by the system without stopping the device. The system now takes the mutated code, and the inputs are provided by the MQTT queue like it would under a normal environment. The outputs of these inputs are captured and saved to see them impact of the mutated code on the results.

*4) The Testing Framework*

The framework includes test derivation, deployment, execution and analysis. Tests are derived considering the IoT specification machines and their corresponding mutants. The tests are then deployed via MQTT(Node-red) on the ESP32. The response from the devices in respect to the inputs of the test cases are captured again via the MQTT and stored in the SQL database. Tests are run (test case runner) one after the other and the corresponding output sequences are recorded and consequently verdicts issues whether the test kills the considered mutant or not by comparing the actual output response of the implementation to the expected response of the test case. The process repeats considering the mutants one after the other.

### a) Tests for detecting security attacks

In general, deriving a test that can distinguish a given state machine from a given (attack) mutant can be carried out using the traditional methods elaborated in [43][44]. However, in our case, as the number of mutants is rather small, we derive these tests by hand. That is, for each attack type, we consider all mutants in that category one after the other; then for each mutant, we derive include in the test suite a test that kills the mutants. The input sequence of the test is then run against all other unconsidered yet mutants to identify those that can also be killed by the test. This reduces the number of derived tests. Here we include an example of such test cases using the motion sensor.

> Test for attack $A_3$: Consider the scenario for attack $A_3$ mentioned in Table I. A mutant $M_{A3}$ of the state machine in Fig. 3 is created such that the output of transition $t_{11}$ is changed to $o_6$ at $S_5$. The test case $(i_5, t=1)/\varepsilon \ (i_5, t=2)/o_6$ detects such a fault as the expected behavior according to the specification is $(i_5, t=1)/\varepsilon \ (i_5, t=2)/o_5$. That is after applying the input $i_5$ at $t=1$ then applying again $i_5$ at $t=2$, the mutant responds to the second input with the output $o_6$ while the expected output is $o_5$.

### b) Tests for detecting actual device manipulation

The test cases remain the same for testing the physical device. For example, the attacker can just remove the edge device and replace it with its own. In this situation, the system will behave normally. We can test this suspicious device by our test cases that we have created for all states. However, there is a possibility, that even then the system could pass all the test cases correctly. The mechanism can be reverse engineered by injecting faults/mutants in the suspicious device. We already know how a device behaves when subjected to different types of faults and we have those results. With these results, how would the suspicious device behave when subjected to those exact same faults. Is the behavior of the suspicious device same as compared to a normal device under a faulty environment? Does it break in the same way that the normal device would break? Does it produce the same kind of error that a normal device had produced when it was tested with these faults? Hence detecting the physical manipulation cannot just be verified with the correct behavior of the system, but it can also be detected by the incorrect behavior

### c) Tests for detecting code mutants

These mutants are generated using a tool that creates changes to the code in different ways (changing the assignment operators, changing the variable names, variable value, etc.). To generate code mutants, the program written for the workflow shown in Fig. 2 is fed into a mutant tool generator. The mutant tool generator creates different files with one mutant each. For Example: If the program has a line written as String msg1 = "Health Check", then its mutated file would change it to String msg1 = "&Health Check".

*5) Implementation tools*

The framework explained in the above sections has been implemented with different interfaces and tools as shown in Table VI.

TABLE VI
IMPLEMENTATION TOOLS AND PROCESS AUTOMATION

| No | Implementation steps | Tools Used |
|---|---|---|
| 1 | Generating the mutants | Python and C++ |
| 2 | Injecting it on the embedded system Over The air | Python scripts and Arduino- |
| 3 | Running the mutated code on the system | Arduino-cli |
| 4 | Capturing the behavior of the system for different inputs/outputs | MQTT, Node-red, sqlite scripts |
| 5 | Running the test cases on these values captured | Gtest and C++ |
| 6 | Displaying the test result | C++ |
| 7 | Automating Step 2-6 | Bash Script |

## VI. EVALUATION

We evaluate the proposed work from several perspectives.

### A. Identify the coverage of security faults

The proposed approach worked very well and resulted in a perfect mutation score of 100% of security attack code mutants. This performance can be compared to an alternative approach that does not consider the type of security threat, but is based on known code-based mutants. For the motion sensor implementation, 40 code-based mutants using the automatically generated mutation operators from a tool [41] were used. Each of these mutants was then injected into the real hardware, and then the 12 security test cases are run against the mutant. The corresponding mutation score was 22.5 percent. For the ultrasonic motion sensor, 40 code mutants are considered and the corresponding 12 test cases obtained a 20 percent mutation score. For the RFID 12 test cases on the 40 mutants, the mutation score was 15 percent. Thus, on average, 19 percent mutation score was obtained. This gives us an idea of the general fault coverage of security tests. In other words, in general, our considered security faults do not correspond to arbitrary mutants (faults) derived using the traditional mutation C testing operators; and thus, there is a need to consider specific tests for these attacks as we do in this paper.

### B. Assess impact of threats in battery drainage:

We conducted an experiment to assess the severity of battery drainage attacks on the battery performance. A big factor contributing to draining is if we make the device perform power consuming sub routines at shorter intervals and more often. We measured the power consumption of the device by injecting mutants that lead to battery drainage using an inline hardware device. The power consumption was calculated using 2 scenarios 1) When the system was running under normal conditions 2) when the system was running under a battery draining attack. The experiment ran for 30 minutes for each device collected at a frequency of every 3 seconds. The results are shown in Table VI.

TABLE VI
POWER CONSUMPTION AND COMPARISON FOR IoT DEVICES UNDER ATTACK

| Device | Energy (in watt-hour) normal conditions | Energy (in watt-hour) under battery draining attack | % increase in energy consumption |
|---|---|---|---|
| Motion Sensor | 900 Watt-hour | 1740 Watt-hour | 48% |
| RFID Card Reader | 1674 watt-hour | 2909 Watt-hour | 42% |
| Ultrasonic motion sensor | 1152 watt-hour | 2070 Watt-hour | 44% |

As Table VI shows, there was a 48% increase in the energy consumption for motion sensor and 42% and 44% for RFID card reader and ultrasonic motion sensor respectively. Such an increase can be controlled if battery drainage attacks can be spotted in real time.

### C. Feasibility of applying the work in practice

As the number of security tests per attack type is rather small, then it is reasonable for assessing a certain anomaly, to run related tests many times as needed without causing battery drainage. We conduct a simple experiment to assess the cost of running the tests on battery drainage. We measured the power consumption by running all the three edge devices for 30 minutes each (motion sensor, RFID card reader and ultrasonic motion sensor) using an inline hardware device to calculate the current. The power consumption was calculated using 2 scenarios 1) When the system was running under normal conditions and performing normal tasks 2) When the system was subjected to continuous testing – This means that the test cases ran on the system continuously for 30 minutes and the power consumption was calculated. The results are shown in Table VII.

TABLE VII
POWER CONSUMPTION AND COMPARISON FOR IoT DEVICES WHEN RUNNING TEST CASES

| Device | Energy (in watt-hour) normal conditions | Energy (in watt-hour) when the tests are continuously running | % Increase in the energy consumption |
|---|---|---|---|
| Motion Sensor | 900 Watt-hour | 1170 Watt-hour | 23% |
| RFID Card Reader | 1674 watt-hour | 2106 Watt-hour | 20.5% |
| Ultraso | 1152 watt-hour | 1368 Watt-hour | 15.7% |

As Table VII shows the percentage of energy consumption was below 25% for all the 3 devices when the tests were being run continuously without a break. Additional modifications can be made to run the tests at regular intervals instead of running them continuously if the power consumption needs to be reduced further.

*D. Applicability of the work to realistic systems*

In this regard, experiments were conducted to determine if the security tests derived by our method could detect the considered anomalies in real implementations. In total, 29, 35, 32 attack mutants were derived for motion sensor, ultra-sonic motion sensor, and RFID, respectively. Each of these anomaly mutants was injected into the real hardware and our experiments showed that each of these anomalies was detected when running the corresponding test case on the real hardware.

## VII. OTHER RELATED WORK

Apart from other techniques that have been used to create intrusion detection systems in IoT devices reported in the Introduction, there has been some work on modelling IoT devices for intrusion detection as well. Fu et al. [53] describe an automata-based intrusion detection system for IoT devices. While they use automata models to automatically detect jam-attack, false-attack, and reply-attack we have used a finite state machine to detect attacks like battery draining, sleep deprivation, data falsification, replay attack and man in the middle attack. Arrington et al. [54] use behavioral anomalies to create an intrusion detection system for IoT devices with behavioral modelling. Their method captures the sensor data as a set of events to give a numerical representation for behavioral identity. They have created NPC (Nonplaying characters) with various roles (owner, attacker, etc). The behavioral patterns are created based on the behavior of these NPCs. The results and conclusion of this method is through simulation whereas we have made the use of the devices in real time, injected mutants and created intrusion in real time rather than a simulated environment. Sedjelmaci et al. [55] used game theory to create a game model of the IoT device for a normal user and an intruder. This technique was used to detect intrusion into a system. The anomaly detection gets activated only when new attack is most likely to be expected. The attack signatures are recorded based on a learning algorithm that builds a rule based on newly attacked pattern it discovers during learning. This was then used to extract the Nash equilibrium value which was in turn used to decide the use of the intrusion detection system for that scenario. Ge et al. [56] used a graphical security model to detect vulnerabilities in cross protocol devices by grouping the IoT devices with same communication protocol They used a multi-layer hierarchical attack representation models by combining attack graphs and attack. They have tried to detect the node controlling and sinkhole attack with this approach and they have also tried to analyze the denial of service attacks using the same approach. While this is also a model-based approach, our methods focus on different types of attacks. Loise et al. [57] introduced 15 general security aware mutation operators for Java programs, different than those we introduced here for edge devices, and they assessed these operators in respect to traditional Java operators.

Thus, our work complements previous related work by considering for the first-time modeling of IoT devices and attacks using the timed finite and extended finite state machine models. In addition, we used the models to create dedicated threat mutants and derive tests that kill these mutants. The dedicated tests are few in numbers and this enables the use of the proposed work in practice in contrast to deriving tests using arbitrary mutation operators or deriving complete tests. [47][58] for detecting arbitrary faults in the implementation under test.

## VIII. CONCLUSION

In this paper, we modeled the behavior of the motion sensors and RFID reader IoT edge devices as finite state machines with timeouts. We also modeled common IoT attacks such as the battery draining, sleep deprivation, data falsification, replay, and man in the middle IoT as special mutants of these machines. Mutation testing is then used to derive tests that distinguish these threats from the common behavior of the machines. The behavior of these mutants is tested in real environment by running the tests on them. We also consider tests for detecting actual physical device manipulation. We implemented and assessed the work on a real environment. The environment includes related hardware architecture, an IoT framework for running a program on microcontroller, data collection for observing the behavior of the program and a testing framework for the derivation and injection of mutants/faults into the program and for detection of these faults in practice. The impact of threats on battery drainage is assessed and it is shown to be high, on average it yields 44.6 percent increase in power consumption. However, according to our experiments, the security tests obtained by our method are few in numbers and thus can be run many times with limited overhead, on average we obtained 25 percent energy consumption overhead when continuously running tests run on the system. Another experiment showed that on average security tests have only 18 percent coverage of arbitrary code-based faults/mutants showing that security faults do not correspond to arbitrary code faults; and thus there is a need to consider the tests we proposed in this paper for detecting IoT threats.

As a part of our future work, we would like to build a stronger test suite that is able to detect more security-based attacks; i.e., consider more types of attacks. Another interesting direction is to assess the work on more IoT devices and conduct relevant case studies. Another area for future work is creating a digital twin of the edge device to get more information on every state. Since the entire workflow is modelled and implemented as a finite state machine, every step in the code can be considered as a state. Whenever a test case fails, the digital twin should be able to give more information by pinging the actual device and gathering data at every state. This can help in localizing the exact place of the fault occurrence.

XXX-X-XXXX-XXXX-X/XX/$XX.00 ©20XX IEEE 14